\documentclass[11pt,a4paper]{article}
\usepackage[T1]{fontenc}
\usepackage[utf8]{inputenc}
\usepackage{lmodern}
\usepackage[margin=2.8cm]{geometry}
\usepackage{amsmath,amssymb,amsthm,bm,mathtools}
\usepackage{microtype}
\usepackage[numbers]{natbib}
\usepackage{lineno}
\usepackage[breaklinks]{hyperref}
\usepackage{booktabs}
\numberwithin{equation}{section}
\newtheorem{proposition}{Proposition}[section]
\newtheorem{corollary}[proposition]{Corollary}
\theoremstyle{remark}
\newcommand{\E}{\mathbb E}
\newcommand{\Cov}{\operatorname{Cov}}
\newcommand{\Var}{\operatorname{Var}}
\newcommand{\tr}{\operatorname{tr}}
\newcommand{\R}{\mathbb R}
\title{Predicting fitness from genomic offset: a multivariate latent-trait framework}
\author{Olivier Fran\c cois}
\date{\today}

\begin{document}

\maketitle

\begin{abstract}
\noindent Genomic offset methods quantify the genetic consequences of environmental change using genotype–environment associations (GEA), but the conditions under which they predict fitness remain poorly understood. Here, we develop a theoretical framework linking genomic offset to relative fitness loss through multivariate latent adaptive traits under stabilizing selection. We show that squared genomic offset is linearly related to expected log-relative-fitness loss when averaged over genetic architectures, provided that direct environmental effects are absent. However, this relationship does not necessarily hold for a realized genetic architecture.
We identify two complementary mechanisms that can explain high predictive accuracy of genomic offset statistics. First, when effective adaptive dimensionality is sufficiently large, the genetically mediated fitness response concentrates around its marginal expectation, reducing the dependence on the particular genetic architecture. Second, even with few adaptive traits, accurate predictions can arise from geometric alignment between the complete fitness-loss matrix and the GEA metric. This second mechanism also encompasses direct environmental effects and their interactions with genetically mediated responses, which may preserve or weaken predictive accuracy depending on their alignment. We further derive bounds on predictive accuracy that characterize these effects.
Our results clarify the assumptions and limitations underlying genomic offset predictions and suggest that genomic offset is best interpreted as a measure of genetically informed adaptive mismatch, rather than a direct predictor of realized fitness or any particular phenotypic trait.
\end{abstract}

\vspace{2cm}

\noindent \textbf{Keywords.} Genomic offset; genotype–environment associations; local adaptation; relative fitness; latent adaptive traits; stabilizing selection; predictive accuracy; genetic architecture.

\vspace{1cm}

\noindent  \textbf{Author's affiliation}:

\vspace{.5cm}

\noindent 
Universit\'e Grenoble-Alpes,  Centre National de la Recherche Scientifique, Grenoble INP, TIMC UMR 5525, 38000 Grenoble, France.

\clearpage
\newpage


\section{Introduction}\label{sec:introduction}
Genotype--environment association (GEA) models describe how genetic variation covaries with environmental gradients and provide a widely used framework for investigating local adaptation \citep{Forester2018}. Their application to genomic offset (GO) extends this descriptive role to prediction: changes in environmental conditions are translated into distances between predicted genetic compositions, which are then interpreted as indicators of potential maladaptation \citep{Fitzpatrick2015,Rellstab2016}. Yet a genomic distance is not, by itself, a measure of fitness. Understanding when GO can predict fitness loss requires an explicit connection between environmental associations at genetic loci, the traits on which selection acts, and the geometry of the fitness response.

Simulation studies and common-garden experiments provide encouraging but heterogeneous evidence. Genomic offset has been compared with simulated fitness or maladaptation under spatially explicit evolutionary scenarios \citep{Laruson2022,Lind2024,Gain2023,Capblancq2025}, and associations with performance have been examined in provenance experiments [\citealp{Rhone2020,Fitzpatrick2021}, Lachmuth {\it et al.} 2023]. However, predictive success varies with population structure, demographic history, the genomic offset method, the genetic and environmental variables retained, and the validation setting. These results have prompted questions about the interpretation of genomic vulnerability and the conditions under which a genomic mismatch can be translated into an expected fitness response \citep{Rellstab2021,Lotterhos2024,Ahrens2025}.

A quantitative theory of GO relates the covariance of GEA coefficients to expected log-relative-fitness loss under stabilizing selection \citep{Gain2023,Capblancq2025}. A central difficulty remains: fitness depends on genetic effects on adaptive traits, which are typically unobserved, whereas GO is constructed from GEA coefficients alone. Moreover, the fitness response of a particular genetic architecture need not coincide with its expectation across hypothetical architectures. This distinction is especially important when adaptation involves several correlated traits \citep{Lande1979, Lande1983, Arnold2001}, or when the environment affects trait mismatch through both genetically mediated and direct pathways.

Here we develop a multivariate latent-trait framework to distinguish the marginal relationship between GO and fitness loss from prediction conditional on a realized genetic architecture. We show that increasing effective adaptive dimensionality can make the fitness-loss matrix concentrate around a scaled version of the GEA metric, thereby improving the reliability of genomic offset predictions. We also derive spectral bounds showing that accurate prediction can arise with only a few adaptive traits, provided that the realized fitness response is sufficiently aligned with the dominant environmental directions of the GEA metric. Finally, we characterize how direct environmental effects and their interactions with genetically mediated effects modify predictive accuracy.

Together, these results show that genomic offset can achieve high predictive accuracy through distinct mechanisms operating under very different adaptive architectures. Consequently, a strong association between genomic offset and fitness loss does not, by itself, reveal the effective dimensionality of adaptation or identify the underlying adaptive traits. Our framework thus clarifies not only when genomic offset can be predictive, but also what can—and cannot—be inferred biologically from its predictive success.

\section{Mathematical framework and modeling assumptions}\label{sec:model}
\subsection{Environmental displacement and genomic offset}\label{subsec:offset}
We consider a population exposed to a change from a reference environment $\bm{x}$ to a new environment $\bm{x}^{\star}$, both in $\R^d$. Following the convention used throughout this manuscript, the environmental displacement is $\bm h=\bm x-\bm x^{\star}$. Reversing this convention changes the signs of the linear mismatch components but not the quadratic quantities considered below, provided that the convention is used consistently. The $d$ environmental coordinates may represent measured bioclimatic variables or transformed environmental gradients.

Let $L$ denote the number of loci contributing to genotype--environment associations (GEA). For each locus $\ell$, let $\bm b_\ell\in\R^d$ be its vector of GEA coefficients, assumed to be corrected for population structure or other confounding factors. Conditional on these coefficients, we define the uncentered GEA second-moment matrix and squared genomic offset by
\begin{equation}\label{eq:offset}
 C_b=\frac1L\sum_{\ell=1}^L\bm b_\ell\bm b_\ell^\top,
 \qquad g(\bm h)=G^2(\bm h)=\bm h^\top C_b\bm h
 =\frac1L\sum_{\ell=1}^L(\bm b_\ell^\top\bm h)^2.
\end{equation}
We refer to $C_b$ as the \emph{GEA metric} matrix, since it defines the quadratic form used to measure environmental displacement in genomic offset, with greater weight assigned to environmental directions associated with stronger multilocus genetic responses \citep{Gain2023}.
The matrix $C_b$ is positive semidefinite and is not, in general, a centered covariance matrix. Genomic offset measures environmental displacement in the metric induced by GEA coefficients; it does not, by itself, specify a fitness response.

\subsection{Latent adaptive traits and genetic architecture}\label{subsec:latent}

We assume that adaptation involves $K$ potentially correlated latent quantitative traits, representing the underlying phenotypic dimensions through which genetic variation and environmental change affect fitness. These traits are not necessarily identifiable with particular measured phenotypes. They rather provide a multivariate representation of adaptive mismatch relative to the local optimum. Their coordinates need not be statistically independent.

For each locus $\ell$, let $\bm a_\ell\in\mathbb R^K$ denote its vector of effects on these traits. We model the genetically mediated component of their response to environmental displacement $\boldsymbol h$ as

\begin{equation}\label{eq:B}
 \bm U(\bm h)=B^\top\bm h,\qquad
 B=\frac1{\sqrt L}\sum_{\ell=1}^L\bm b_\ell\bm a_\ell^\top
 \in\R^{d\times K}.
\end{equation}
The $L^{-1/2}$ normalization gives a nondegenerate genetic-response covariance as the number of independent loci grows, under the moment assumptions below. 

Throughout, the GEA coefficient vectors $\bm b=(\bm b_1,\ldots,\bm b_L)$ are treated as fixed when studying uncertainty in the genetic architecture.  Expectations and probabilities indexed by $a$ are therefore conditional on $\bm b$, and this conditioning is omitted from the notation. We assume that the allelic-effect vectors are independent across loci and satisfy
\begin{equation}\label{eq:allelic}
 \E_a[\bm a_\ell ]=\bm0,\qquad
 \Cov_a(\bm a_\ell )=\Sigma_a,
 \qquad \ell=1,\ldots,L,
\end{equation}
where $\Sigma_a\in\R^{K\times K}$ is symmetric positive semidefinite and common to all loci.

The centering assumption is an ensemble assumption about possible genetic architectures, not a statement that the effects of loci in a particular organism are zero. 
In a joint evolutionary model, local adaptation could induce conditional means or covariance structures of allelic effects that depend on the observed GEA coefficients \citep{Yeaman2015, Lasky2023}. Such dependencies are not explicitly modeled here, since our framework specifies the conditional distribution of allelic effects for fixed GEA coefficients.
In addition, these assumptions should not be interpreted as a generative model of genetic architecture, but rather as part of a probabilistic framework representing uncertainty about unobserved genetic effects. This distinction motivates the subsequent analyses of marginal and realized fitness predictions.

Equations~\eqref{eq:offset}--\eqref{eq:allelic} imply the exact identities
\begin{equation}\label{eq:covU}
 \E_a[\bm U(\bm h) ]=\bm0,\qquad
 \Cov_a\bigl(\bm U(\bm h) \bigr)
 =g(\bm h)\Sigma_a.
\end{equation}
 When Gaussian allelic effects are explicitly assumed, $B$ has a matrix-normal distribution with row covariance $C_b$ and column covariance $\Sigma_a$ (possibly degenerate). No Gaussian assumption is needed for the covariance identity. The Gaussian assumption will be invoked only for distributional and concentration results.

\subsection{Multivariate stabilizing selection and direct environmental effects}\label{subsec:fitness}
Let $\bm Z(\bm h)\in\R^K$ denote the deviation of the latent trait vector from its local adaptive optimum. We decompose this mismatch into genetically mediated, direct environmental, and residual components:
\begin{equation}\label{eq:Z}
 \bm Z(\bm h)=B^\top\bm h+C^\top\bm h+\bm\varepsilon.
\end{equation}
Here $C \in \mathbb{R}^{d \times K}$ is a fixed matrix describing direct environmental contributions to mismatch, including pathways not mediated by the modeled GEA effects. Such contributions may reflect phenotypic plasticity and other environmentally induced changes in adaptive traits \citep{Pigliucci2001}. The residual vector $\bm\varepsilon\in\R^K$ is assumed independent of the allelic effects, with $\E[\bm\varepsilon]=\bm0$ and $\Cov(\bm\varepsilon)=\Sigma_\varepsilon$. We assume that its distribution and covariance do not depend on $\bm h$; environmental heteroscedasticity would require a corresponding extension. 

Local relative fitness is evaluated in the destination environment $x^\star$, relative to the maximum fitness attained by a population whose adaptive traits match the local optimum. Under multivariate Gaussian stabilizing selection, it is modeled as
\begin{equation}\label{eq:omega}
 \omega(\bm h)=\exp\!\left[-\frac12\bm Z(\bm h)^\top S\bm Z(\bm h)\right],
\end{equation}
where $S\in\R^{K\times K}$ is symmetric positive semidefinite ($S=S^\top\succeq0$), and encodes the strengths and joint directions of stabilizing selection \citep{Kirkpatrick2009, WalshLynch2018}. The matrix $S$ is held fixed across environmental displacements. 

The primary response analyzed in this study is the negative logarithm of relative fitness
\begin{equation}\label{eq:Q}
 Q(\bm h)=-\log\omega(\bm h)
 =\frac12\bm Z(\bm h)^\top S\bm Z(\bm h),
\end{equation}
which quantifies fitness loss on a logarithmic scale. Thus $Q$ is nonnegative, and larger values indicate greater maladaptation relative to the optimum in the destination environment. Relative fitness is distinguished from absolute fitness: an environment-dependent maximum absolute fitness, if present, is not modeled by $Q$.

\subsection{Marginal (annealed) and realized (quenched) prediction}\label{subsec:annealed-quenched}
Two different averaging operations must be distinguished. The \emph{marginal} perspective averages over possible genetic architectures, whereas the realized or \emph{quenched} perspective evaluates prediction for a single, fixed genetic architecture, as encountered in a particular population. In the marginal or annealed model, allelic effects are regarded as random and both genetic-architecture uncertainty and residual variation are integrated out. Conditional on the GEA coefficients and the environmental displacement, the mean log-fitness loss is
\begin{equation}\label{eq:mAdef}
 m_{\mathrm A}(\bm h)
 \stackrel{\mathrm{def}}{=}\E_{a,\varepsilon}\bigl[Q(\bm h)\mid\bm h  \bigr].
\end{equation}
This quantity characterizes an ensemble of possible genetic architectures consistent with the assumed allelic-effect distribution. It is not necessarily the expected fitness response of one particular population.

In the \emph{realized} or \emph{quenched} model, the matrix $B$ is held fixed at its realized value, even if its entries are unknown to the investigator. Only residual variation is averaged out:
\begin{equation}\label{eq:mQdef}
 m_{\mathrm Q}(\bm h;B)
 \stackrel{\mathrm{def}}{=}\E_\varepsilon\bigl[Q(\bm h)\mid\bm h,B  \bigr].
\end{equation}
With total effect $T=B+C$, the quenched mean takes the quadratic form
\begin{equation}\label{eq:mQ}
 m_{\mathrm Q}(\bm h;B)
 =\alpha+\frac12\bm h^\top M_B\bm h,
 \qquad \alpha=\frac12\tr(S\Sigma_\varepsilon),\qquad
 M_B=TST^\top.
\end{equation}
The matrix $M_B$, called the \emph{realized fitness matrix}, is positive semidefinite, but generally differs from a scalar multiple of the GEA metric, $C_b$. Its expansion,
\begin{equation}\label{eq:cross}
 M_B=BSB^\top+CSC^\top+BSC^\top+CSB^\top,
\end{equation}
separates genetically mediated, direct, and genetic--direct interaction terms. Centering the allelic effects makes the interaction terms vanish after marginal averaging, but does not eliminate them for a fixed architecture.

By the law of iterated expectation, $m_{\mathrm A}(\bm h)=\E_a[m_{\mathrm Q}(\bm h;B) ]$. More explicitly, the covariance identity~\eqref{eq:covU} yields
\begin{equation}\label{eq:mA}
 m_{\mathrm A}(\bm h)
 =\alpha+\gamma g(\bm h)
 +\frac12\bm h^\top CSC^\top\bm h,
 \qquad \gamma=\frac12\tr(S\Sigma_a).
\end{equation}
In particular, when direct environmental effects are absent, the marginal mean loss is exactly affine in squared genomic offset. This equality of expectations does not imply that the quenched response is affine for a particular realized architecture. Later sections establish when and how the realized fitness matrix can approach the GEA metric.

\subsection{Scope of fitness predictions}\label{subsec:scope}
Our theoretical comparisons concern the conditional mean of $Q=-\log\omega$, not necessarily the log of mean fitness. In general, the two quantities differ
\begin{equation}\label{eq:jensen}
 \E[Q\mid\bm h]\ne-\log\E[\omega\mid\bm h],
\end{equation}
with the difference arising from variation in latent trait mismatch. Similarly, observed individual fitness losses fluctuate around their conditional mean because of residual variation; marginal predictions additionally average over genetic architectures. Consequently, prediction of mean log-relative fitness loss, prediction of realized mean log-relative fitness loss, and prediction of individual fitness observations are separate statistical questions. The results below first address the mean relationships and subsequently quantify their variability and predictive accuracy.

\section{Main results}\label{sec:results}
Our results distinguish two questions that are often conflated in genomic offset applications. First, why should the genomic offset predict maladaptation on average over possible genetic architectures? Second, under what conditions does this prediction remain accurate for the fixed, generally unknown architecture of a population? The marginal identity in Eq.~\eqref{eq:mA} answers the first question. The following results address the second question, first through adaptive dimensionality and matrix concentration, and then through the geometry of environmental variation and direct environmental effects. Proofs are collected in Section~\ref{sec:proofs}.

\subsection{The role of effective adaptive dimensionality}\label{subsec:keff}

Previous studies in quantitative genetics have emphasized that the response to selection may be constrained to a limited number of effective dimensions, even when many traits are considered \citep{Kirkpatrick2009, HansenHoule2008}. These studies primarily concern genetic variation and evolutionary responses to selection, rather than fitness loss. Here, we introduce a related but distinct concept of \emph{effective adaptive dimensionality}, which accounts for both the covariance of allelic effects and the strength and structure of stabilizing selection, and determines the concentration of genetically mediated fitness loss around its marginal expectation.

In the absence of direct and residual effects ($C=0$, $\Sigma_\varepsilon=0$), the marginal mean of log-relative fitness loss is $\E_a[Q\mid\bm h  ]=\gamma g(\bm h)$. 
However, fitness loss may vary substantially across genetic architectures, even for a fixed environmental displacement. To quantify this variation relative to the marginal mean, we define
\begin{equation}\label{eq:keff}
 K_{\mathrm{eff}}=\frac{[\tr(S\Sigma_a)]^2}{\tr(S\Sigma_a S\Sigma_a)},
\end{equation}
when $\tr(S\Sigma_a)>0$. This quantity is the participation ratio of the eigenvalues of $S^{1/2}\Sigma_a S^{1/2}$; it lies between one and the rank of this matrix, and hence does not exceed $K$.
 In our framework, effective adaptive dimensionality accounts for the strength and structure of stabilizing selection and determines the concentration of genetically mediated fitness loss around its marginal expectation.

\begin{proposition}[Relative variation and effective adaptive dimensionality]\label{prop:keff}
Suppose the allelic effects are conditionally independent, centered and Gaussian with common covariance $\Sigma_a$, and $C=0$, $\Sigma_\varepsilon=0$. For $g(\bm h)>0$ and $\gamma>0$, the squared coefficient of variation of the negative log-relative fitness is
\begin{equation}\label{eq:cv}
 \frac{\Var_a(Q(\bm h)\mid\bm h  )}{\E_a[Q(\bm h)\mid\bm h  ]^2}
 =\frac{2}{K_{\mathrm{eff}}}.
\end{equation}
\end{proposition}
\paragraph{}
The number of loci $L$ does not appear in Eq.~\eqref{eq:cv}. Even with very many loci, a small number of effectively independent, selection-weighted adaptive traits can produce substantial differences between realized fitness loss and its marginal expectation. Genetic correlations and unequal selection strengths reduce $K_{\mathrm{eff}}$: counting measured or hypothesized traits is therefore not sufficient to assess predictive reliability. The Gaussian assumption is needed for this exact variance formula, whereas the marginal mean identity (2.12) only requires the specified first and second moments of allelic effects. This result concerns variation over architectures at a fixed displacement, rather than correlation across environments for one architecture.

\paragraph{Example.} Nominal versus effective adaptive dimensionality.  Consider a population whose adaptation to environmental change depends on $K$ quantitative traits, such as physiological, morphological, or life-history traits. For simplicity, assume that all traits experience stabilizing selection of equal strength, with no interactions between trait deviations in the fitness function ($S=I_K$, in appropriately scaled trait coordinates). Genetic effects on different traits are independent, but their variances may differ, reflecting unequal contributions to the genetically mediated response to environmental change. We represent these differences by $\Sigma_a=\operatorname{diag}(v_1,\ldots,v_K)$, where $v_k$ is the variance of allelic effects on trait $k$. We focus on the genetically mediated component of log-fitness loss, independently of direct environmental effects.
Under these assumptions, the effective adaptive dimensionality is

$$
K_{\mathrm{eff}}=
\frac{\left(\sum_{k=1}^K v_k\right)^2}
{\sum_{k=1}^K v_k^2}.
$$

If all traits contribute equally ($v_1=\cdots=v_K=1$), then $K_{\mathrm{eff}}=K$, and the relative variance of genetically mediated log-fitness loss across genetic architectures is $2/K$. Thus, increasing the number of equally contributing traits makes the realized response increasingly concentrated around its marginal expectation.

In contrast, suppose that one trait dominates the genetic response, with $v_1=1$ and $v_2=\cdots=v_K=v$, where $0<v<1$. Then

$$
K_{\mathrm{eff}}=
\frac{[1+(K-1)v]^2}
{1+(K-1)v^2}.
$$

As $v\to0$, $K_{\mathrm{eff}}\to 1$, regardless of the nominal number of traits. Consequently, the relative variance approaches 2, and concentration is lost. This example illustrates that predictability through concentration depends on the number of effectively contributing adaptive traits rather than on the nominal number of traits or genetic loci.

\subsection{Concentration of the genetically mediated fitness matrix}\label{subsec:matrix}
The realized genetic contribution to fitness loss is governed by the matrix $BSB^\top$, whereas genomic offset is defined by the GEA metric $C_b$. For genomic offset to predict genetic fitness loss across environmental displacements, the normalized matrix $BSB^\top/(2\gamma)$ should be close to $C_b$.

\begin{proposition}[Concentration of the genetic fitness matrix]\label{prop:matrix}
Under the Gaussian allelic-effect assumptions of Proposition~\ref{prop:keff}, conditionally on the GEA coefficients,
\begin{equation}\label{eq:matrixmean}
 \E_a[BSB^\top]=2\gamma C_b,
\end{equation}
and
\begin{equation}\label{eq:matrixconcentration}
 \E_a\!\left[\left\|\frac{BSB^\top}{2\gamma}-C_b\right\|_F^2 \right]
 =\frac{\|C_b\|_F^2+[\tr(C_b)]^2}{K_{\mathrm{eff}}}.
\end{equation}
\end{proposition}
\paragraph{}
This identity quantifies the discrepancy between the genetically mediated fitness geometry and the GEA geometry. For fixed environmental dimension and GEA metric, the normalized genetic fitness matrix converges in probability to $C_b$ as $K_{\mathrm{eff}}$ increases. 
After normalization by $\|C_b\|_F^2$, the identity above also shows that relative mean-square matrix concentration occurs if and only if $(1+r_{\mathrm{eff}}(C_b))/K_{\mathrm{eff}}\to 0$, where $r_{\mathrm{eff}}(C_b)=[\operatorname{tr}(C_b)]^2/\operatorname{tr}(C_b^2)$. In particular, for a fixed environmental dimension $d$, this condition is equivalent to $K_{\mathrm{eff}}\to\infty$, providing a necessary and sufficient condition for relative mean-square concentration under the Gaussian model.
 
The concentration result concerns the genetically mediated component $BSB^\top$, irrespective of the presence of direct environmental effects. When $C=0$, however, the complete fitness matrix reduces to $M_B=BSB^\top$. As effective adaptive dimensionality increases, this matrix concentrates around $2\gamma C_b$, so that the quenched mean log-fitness loss approaches its marginal expectation, $m_Q(h;B)\approx\alpha+\gamma g(h)$. Thus, in the absence of direct environmental effects, squared genomic offset becomes an increasingly accurate predictor of mean log-fitness loss for a realized genetic architecture. Biologically, a sufficiently diversified adaptive architecture can average out the idiosyncratic contributions of individual latent traits, making the genetically mediated fitness response increasingly predictable from genomic offset.

\subsection{Geometric limits to genomic offset predictive accuracy}\label{subsec:spectral}
To assess prediction across environmental displacements, suppose $\bm h\sim\mathcal N(0,\Sigma_x)$ with $\Sigma_x$ positive definite. For a fixed realized architecture, predictive accuracy can be defined as the squared correlation, across environmental displacements, between squared genomic offset and conditional mean negative log-relative-fitness loss
\begin{equation}\label{eq:r2_q}
R_{\mathrm Q}^2=\operatorname{Corr}_{\bm h}^2(g(\bm h),m_{\mathrm Q}(\bm h;B)),
\end{equation}
whenever both variances are nonzero. Put
\begin{equation}\label{eq:whiten}
 A=\Sigma_x^{1/2}C_b\Sigma_x^{1/2},\qquad F=\Sigma_x^{1/2}M_B\Sigma_x^{1/2}.
\end{equation}
The matrix $A$ and $F$ represent the GEA metric and realized fitness matrices in environmentally whitened coordinates, accounting for the covariance structure of environmental variation. 
Equivalently, one may first orthogonalize and scale the environmental variables to unit covariance. In these whitened coordinates, the GEA and fitness matrices become $C_b=A$ and $M_B=F$, respectively.
A useful intermediate identity is
\begin{equation}\label{eq:r2}
 R_{\mathrm Q}^2=\frac{[\tr(AF)]^2}{\tr(A^2)\tr(F^2)}.
\end{equation}
Thus, predictive accuracy is the squared Frobenius cosine between the environmentally weighted GEA and fitness-loss matrices. This quantity measures their \emph{spectral alignment}, reflecting both the correspondence between their eigendirections and the relative weights assigned to these directions. In particular, a change in the strength of selection that simply rescales $F$ does not change $R_{\mathrm Q}^2$.

\begin{proposition}[Spectral bounds on predictive accuracy]\label{prop:spectral}
Let $r=\operatorname{rank}(F)\leq\min(K,d)$, where $K$ is the number of latent adaptive traits and $d$ the number of environmental variables. Let $\lambda_1\ge\cdots\ge\lambda_d\ge0$ be the eigenvalues of $A\ne0$. Then
\begin{equation}\label{eq:spectral}
 \frac{\lambda_d^2}{\sum_{i=1}^d\lambda_i^2}\le R_{\mathrm Q}^2  \le\frac{\sum_{i=1}^{r}\lambda_i^2}{\sum_{i=1}^d\lambda_i^2} 
 \le\frac{\sum_{i=1}^{\min(K,d)}\lambda_i^2}{\sum_{i=1}^d\lambda_i^2}.
\end{equation}
\end{proposition}
\paragraph{}  High predictive accuracy can arise even with few latent adaptive traits when the realized fitness response is concentrated along the dominant directions of the GEA metric. The upper bound describes a geometric limitation independent of matrix concentration. If genomic offset varies along many comparably important environmental directions, a fitness response of low rank cannot reproduce all of them. Conversely, when the GEA spectrum is dominated by a few directions, strong prediction is possible even with few adaptive traits. However, the spectrum only determines the maximum achievable predictive accuracy for a given rank of the fitness response; whether this bound is approached depends on the alignment of the relevant environmental directions and the relative weights assigned to them by the GEA and fitness matrices.

\paragraph{Example.} High predictive accuracy with one latent trait.
Consider two independent, standardized environmental variables,
$K=1$, and no direct effects. In whitened coordinates, let

$$
 A=\begin{pmatrix}1&0\\0&\delta\end{pmatrix},
 \qquad
 F=vv^\top,\qquad
 v=\begin{pmatrix}\cos\theta\\\sin\theta\end{pmatrix},
 \qquad 0<\delta\leq1.
$$
By Eq.~(3.7),

$$
 R_Q^2=\frac{(\cos^2\theta+\delta\sin^2\theta)^2}
 {1+\delta^2}.
$$
For $\theta=0$, this gives $R_Q^2=1/(1+\delta^2)$,
which approaches one as $\delta\to0$; for $\theta=\pi/2$,
it gives $R_Q^2=\delta^2/(1+\delta^2)$, which approaches zero.
Thus, even with a single adaptive trait, genomic offset can be
nearly perfectly predictive when the realized fitness response
follows the dominant GEA direction, but poorly predictive when
it follows a weak GEA direction. This is a conditional example
for a realized architecture, not a statement about how frequently
such orientations arise under the allelic-effect model.

\paragraph{Predictive accuracy for observed log-fitness loss.}
The coefficient $R_Q^2$ measures predictive accuracy for the
conditional mean log-fitness loss, rather than for individual
fitness observations. Consider a linear regression of observed
log-fitness losses $Q(h_i)$ on squared genomic offsets
$g(h_i)$, with an intercept. The population coefficient of
determination satisfies
\begin{equation}
R_{\mathrm{obs}}^2
=
\operatorname{Corr}_{\bm h,\varepsilon}^2
\bigl(g(\bm h),Q(\bm h;B)\bigr)
=
R_Q^2
\left(
1+
\frac{
\mathbb E_{\bm h}[
\operatorname{Var}_{\varepsilon}(Q\mid \bm h,B)]
}{
\operatorname{Var}_{\bm h}(m_Q(\bm h;B))
}
\right)^{-1}.
\end{equation}
The multiplicative factor can be interpreted as a
signal-to-total-variance ratio, where the signal is the
variation in conditional mean log-fitness loss across
environmental displacements and the noise is the
expected residual variance. Under Gaussian residual effects
$\varepsilon\sim\mathcal N(0,\Sigma_\varepsilon)$,
independent of environmental displacement, this becomes
\begin{equation}
R_{\mathrm{obs}}^2
=
R_Q^2
\left[
1+
\frac{
\operatorname{tr}[(S\Sigma_\varepsilon)^2]
+
2\operatorname{tr}
(TS\Sigma_\varepsilon ST^\top\Sigma_x)
}{
\operatorname{tr}(F^2)
}
\right]^{-1}
\leq R_Q^2.
\end{equation}
Residual variation in adaptive traits reduces the proportion
of observed log-fitness variance explained by genomic offset,
even when the conditional mean log-fitness loss is accurately
predicted. Consequently, high predictive accuracy for mean
log-fitness loss does not necessarily translate into high
explanatory power in empirical fitness regressions
(see Discussion).

\subsection{Reliability of realized predictions with direct environmental effects}\label{subsec:directmain}
We now combine random fluctuations in genetic architecture with deterministic direct effects. Write $U=\Sigma_x^{1/2}B$, $D=\Sigma_x^{1/2}C$, and
\begin{equation}\label{eq:FX}
 F=(U+D)S(U+D)^\top=USU^\top+J+X,\quad
 J=DSD^\top,\quad X=USD^\top+DSU^\top.
\end{equation}
Here $F$ is the environmentally whitened \emph{complete} realized fitness matrix, and $J$
is the environmentally whitened direct-effect matrix. We define the effective rank of the GEA metric as
\begin{equation}\label{eq:reff}
 r_{\mathrm{eff}}(A)=\frac{[\tr(A)]^2}{\tr(A^2)}.
\end{equation}
\begin{proposition}[Probabilistic reliability of realized genomic offset predictions]\label{prop:direct}
Assume centered, independent Gaussian allelic effects with covariance $\Sigma_a$, $\gamma>0$, and $A\ne0$. Then $\E_a[F]=2\gamma A+J$, $\E_a[X]=0$, and
\begin{align}\label{eq:crossvariance}
 \E_a[\|X\|_F^2] &=2\tr(A)\tr(SD^\top DS\Sigma_a)
 +2\tr(ADS\Sigma_a SD^\top).
\end{align}
Define
\begin{equation}\label{eq:eta}
 \eta_{\mathrm{tot}}=\frac{1+r_{\mathrm{eff}}(A)}{K_{\mathrm{eff}}}
 +\frac{\|J\|_F^2+\E_a[\|X\|_F^2]}{4\gamma^2\|A\|_F^2}.
\end{equation}
Then
\begin{equation}\label{eq:directmoment}
 \E_a\!\left[\left\|\frac{F}{2\gamma}-A\right\|_F^2 \!\right]
 =\eta_{\mathrm{tot}}\|A\|_F^2 \, ,
\end{equation}
and for every $0<t<1$,
\begin{equation}\label{eq:directprob}
 \mathbb{P}_a\!\left(1-R_{\mathrm Q}^2\le\frac{t^2}{(1-t)^2}\right)
 \ge 1-\frac{\eta_{\mathrm{tot}}}{t^2}.
\end{equation}
Consequently, $\eta_{\mathrm{tot}}\to0$ implies $R_{\mathrm Q}^2\to1$ in probability.
\end{proposition}

\paragraph{The case without direct effects.}
When $C=0$, both $J$ and $X$ vanish and $\eta_{\mathrm{tot}}=(1+r_{\mathrm{eff}}(A))/K_{\mathrm{eff}}$. Thus a large effective adaptive dimensionality relative to the effective environmental rank is sufficient for high quenched predictive accuracy. This is a sufficient, not necessary, condition; the spectral result shows that strong alignment can also yield high accuracy for low-rank architectures.

\paragraph{Geometric alignment of direct environmental effects.}
The bound above measures the full magnitude of the environmentally whitened direct-effect matrix, $J$, and is deliberately conservative. For $J\ne0$, we define the \emph{geometric alignment of direct environmental effects with the GEA metric} as the Frobenius cosine between their environmentally whitened quadratic-form matrices, $A$ and $J$. This quantity, denoted by $\rho_D$, equals the correlation between squared genomic offset and the direct contribution to fitness loss across Gaussian environmental displacements
\begin{equation}\label{eq:rhoD}
 \rho_D=\frac{\tr(AJ)}{\|A\|_F\|J\|_F}.
\end{equation}
Perfect alignment ($\rho_D=1$) implies that direct effects are proportional to the GEA metric, whereas zero alignment ($\rho_D=0$) indicates orthogonal quadratic forms.

Since $A,J\succeq0$, $0\le\rho_D\le1$. Decompose $J=\rho A+J_\perp$, where $\rho=\tr(AJ)/\|A\|_F^2$ and $\tr(AJ_\perp)=0$. Then
\begin{equation}\label{eq:orthdirect}
 \frac{\|J_\perp\|_F}{2\gamma\|A\|_F}
 =d_D\sqrt{1-\rho_D^2}, \qquad
 d_D=\frac{\|J\|_F}{2\gamma\|A\|_F}.
\end{equation}
This separates the magnitude of direct effects from their mismatch with the GEA geometry. In the marginal response, an exactly aligned direct component changes the slope but not the correlation with genomic offset. 

For a realized architecture, interaction cross terms must also be controlled. Set
\begin{equation}\label{eq:eta_1}
 \eta =\frac{1+r_{\mathrm{eff}}(A)}{K_{\mathrm{eff}}}
 +\frac{\E_a[\|X\|_F^2]}{4\gamma^2\|A\|_F^2}.
\end{equation}

\begin{corollary}[Alignment-aware prediction bound]\label{cor:alignment}
For any $t>0$ such that 
$$
\big( d_D\sqrt{1-\rho_D^2} + t  \big) \stackrel{\mathrm{def}}{=} b <a\stackrel{\mathrm{def}}{=} \big( 1+d_D\rho_D  \big),
$$
we have
\begin{equation}\label{eq:alignmentbound}
 \mathbb{P}_a\left(1-R_{\mathrm Q}^2\le\frac{b^2}{(a-b)^2}\right)
 \ge1-\frac{\eta}{t^2}.
\end{equation}
In particular, if $d_D\sqrt{1-\rho_D^2}\to0$ and $\eta\to0$, then $R_{\mathrm Q}^2\to1$ in probability.
\end{corollary}
\paragraph{}
Direct environmental responses can be large yet compatible with accurate genomic offset prediction when they follow the same environmental geometry as the GEA metric. For example, plastic responses to temperature and water availability might reinforce the composite gradient captured by GEA. Conversely, smaller direct effects acting along poorly aligned environmental directions can reduce predictive accuracy. Thus, the geometry of direct effects can matter more than their absolute magnitude.

Genetic–direct interaction terms introduce an additional source of architecture-specific variation, whose effect on predictive accuracy also depends on geometric alignment. For a realized genetic architecture, the interaction matrix can be decomposed into components parallel and orthogonal to the GEA metric, in the Frobenius sense. The parallel component modifies the part of the fitness response captured by genomic offset, whereas the orthogonal component may introduce variation that reduces predictive accuracy. Interaction terms may also compensate for misalignment arising from the genetic or direct components. Thus, predictive accuracy ultimately depends on the alignment of the complete fitness-loss matrix with the GEA metric, rather than on the separate magnitudes of its components. The probabilistic bounds derived above provide sufficient conditions for accurate prediction, but may be conservative when substantial alignment or cancellation occurs.

\paragraph{Example.} Direct environmental effects and genetic–direct interactions. Consider two environmental variables and two latent adaptive traits ($d=K=2$), with $S=\Sigma_x=I_2$. This example illustrates that the magnitude and alignment of direct effects alone do not determine predictive accuracy. Genetic–direct interactions can modify the geometry of the complete fitness response, and accurate prediction ultimately depends on the alignment of the complete fitness-loss matrix $F$ with the GEA metric matrix $A$. 

For simplicity, assume that the GEA metric and the realized genetically mediated response are given by

$$
A=\begin{pmatrix}1&0\\0&0\end{pmatrix},
\qquad
U=\begin{pmatrix}u&0\\0&0\end{pmatrix},
$$
where $u>0$. Thus, genomic offset depends only on the first environmental direction, and the genetically mediated fitness response is perfectly aligned with the GEA metric. First, consider direct effects aligned with the genetic response,

$$
D=\begin{pmatrix}c&0\\0&0\end{pmatrix}.
$$
The complete fitness matrix is

$$
F=(U+D)(U+D)^\top
=\begin{pmatrix}(u+c)^2&0\\0&0\end{pmatrix}.
$$
Consequently, $R_{\mathrm Q}^2=1$ whenever $u+c\ne0$. Direct effects can therefore be arbitrarily large without reducing predictive accuracy, provided that they preserve the geometry of the GEA metric. The interaction term changes the magnitude of fitness loss but not its environmental direction. 
Now, consider direct effects acting along the second environmental direction. If

$$
D=\begin{pmatrix}0&0\\0&c\end{pmatrix},
$$
then $X=0$ and

$$
R_{\mathrm Q}^2=\frac{u^4}{u^4+c^4}.
$$
Alternatively, if

$$
D=\begin{pmatrix}0&0\\c&0\end{pmatrix},
$$
the direct-effect matrix $J$ is unchanged, but the interaction term is nonzero:

$$
X=\begin{pmatrix}0&uc\\uc&0\end{pmatrix},
\qquad
R_{\mathrm Q}^2=\frac{u^4}{(u^2+c^2)^2}.
$$
Thus, two direct responses with identical quadratic contributions $J$ can lead to different predictive correlations because of their interactions with the realized genetic response. In the latter case, the interaction introduces an additional component orthogonal to the GEA metric in the Frobenius sense.

\section{The single-trait case ($K=1$)}\label{sec:single}
The case of a single latent adaptive trait is particularly informative \cite{Gain2023, Capblancq2025}. It isolates the difference between an exact marginal relationship and prediction for a realized genetic architecture, and shows why increasing the number of loci alone cannot ensure accurate genomic offset predictions. Throughout this section, write $S=\sigma_s^{-2}$, $\Sigma_a=\sigma_a^2$, and $\Sigma_\varepsilon=\sigma_\varepsilon^2$, with $\sigma_s^2>0$. Define the mediated, direct and total effects as follows:
\begin{equation}\label{eq:single-vectors}
 \bm\beta=\frac1{\sqrt L}\sum_{\ell=1}^L a_\ell\bm b_\ell,\qquad C=\bm c, \qquad
 \bm\theta=\bm\beta+\bm c \, ,
\end{equation}
all in $\R^d$. The genetically mediated mismatch is $\bm\beta^\top\bm h$, the direct mismatch is $\bm c^\top\bm h$, and the total mismatch is $\bm\theta^\top\bm h+\varepsilon$. Hence
\begin{equation}\label{eq:single-Q}
 Q(\bm h)=\frac{(\bm\theta^\top\bm h+\varepsilon)^2}{2\sigma_s^2},\qquad
 \omega(\bm h)=\exp[-Q(\bm h)].
\end{equation}

\subsection{Marginal and realized mean fitness loss}\label{subsec:single-means}
The general expressions of marginal and realized mean fitness  simplify to
\begin{equation}
 m_{\mathrm A}(\bm h)
 =\frac{\sigma_\varepsilon^2+\sigma_a^2g(\bm h)+(\bm c^\top\bm h)^2}{2\sigma_s^2},
 \label{eq:single-mA}
\end{equation}
and
\begin{equation}
 m_{\mathrm Q}(\bm h;\bm\beta)
=\frac{\sigma_\varepsilon^2+(\bm\theta^\top\bm h)^2}{2\sigma_s^2}.\label{eq:single-mQ}
\end{equation}
The marginal mean is affine in squared genomic offset when $\bm c=\bm0$, whereas the quenched mean depends on the realized vector $\bm\beta$. Expanding the latter reveals the direct--genetic interaction term:
\begin{equation}\label{eq:single-cross}
 2\sigma_s^2[m_{\mathrm Q}(\bm h;\bm\beta)-\alpha]
 =(\bm\beta^\top\bm h)^2+(\bm c^\top\bm h)^2
 +2(\bm\beta^\top\bm h)(\bm c^\top\bm h),
 \quad \alpha=\frac{\sigma_\varepsilon^2}{2\sigma_s^2}.
\end{equation}
The interaction term vanishes upon averaging over centered allelic effects, but for a fixed architecture it may increase or decrease maladaptation. Biologically, a plastic or other direct environmental response may either reinforce the genetically mediated displacement from the optimum or compensate for it. The marginal expectation alone cannot determine which outcome occurs in a particular population.

\subsection{Variability and the absence of locus-number concentration}\label{subsec:single-variance}
Under Gaussian allelic and residual effects, put
\begin{equation}\label{eq:single-v}
 v(\bm h)=\sigma_a^2g(\bm h)+\sigma_\varepsilon^2.
\end{equation}
The marginal and quenched variances of log-fitness loss are, respectively,
\begin{align}
 \Var_{a,\varepsilon}(Q\mid\bm h  )
 &=\frac{v(\bm h)^2}{2\sigma_s^4}
 +\frac{(\bm c^\top\bm h)^2v(\bm h)}{\sigma_s^4},\label{eq:single-varA}\\
 \Var_\varepsilon(Q\mid\bm h,\bm\beta)
 &=\frac{\sigma_\varepsilon^4}{2\sigma_s^4}
 +\frac{\sigma_\varepsilon^2(\bm\theta^\top\bm h)^2}{\sigma_s^4}.\label{eq:single-varQ}
\end{align}
In the absence of direct $(\bm c = 0)$ and residual effects, we obtain $\gamma=\sigma_a^2/(2\sigma_s^2)$ and
\begin{equation}\label{eq:single-cv}
 \E_a[Q\mid\bm h]=\gamma g(\bm h),\qquad
 \Var_a(Q\mid\bm h)=2\gamma^2g(\bm h)^2,\qquad
 \frac{\Var_a(Q\mid\bm h)}{\E_a[Q\mid\bm h]^2}=2,
\end{equation}
provided $g(\bm h)>0$ and $\sigma_a^2>0$. In this case, we have $K_{\mathrm{eff}}=1$ regardless of $L$. For Gaussian allelic effects, the scaled value of $Q(\bm h)$ has a chi-squared distribution
\begin{equation}\label{eq:single-chi-central}
 \frac{Q(\bm h)}{\gamma g(\bm h)}\sim\chi_1^2.
\end{equation}
Thus increasing the number of independent associated loci does not reduce the relative dispersion of fitness loss across genetic architectures. The central limit theorem makes the aggregate genetic mismatch approximately Gaussian as $L$ grows, but squaring a single Gaussian coordinate does not produce concentration. A large genomic dataset can therefore support a precise estimate of the GEA metric without guaranteeing that this metric closely predicts fitness for the realized genetic architecture.

\subsection{Distribution of log-fitness loss and mean relative fitness}\label{subsec:single-distribution}
Assume in this subsection that allelic and residual effects are Gaussian and independent, as specified in the model. The marginal mismatch is normal with mean $\mu_{\mathrm A}(\bm h)=\bm c^\top\bm h$ and variance $v(\bm h)$. Whenever $v(\bm h)>0$, its squared standardized value has a noncentral chi-square distribution:
\begin{equation}\label{eq:single-noncentral-A}
 \frac{2\sigma_s^2}{v(\bm h)}Q(\bm h)
 \ \overset{\mathrm A}{\sim}\ \chi_1^2\!\left(\lambda_{\mathrm A}(\bm h)\right),\qquad
 \lambda_{\mathrm A}(\bm h)=\frac{(\bm c^\top\bm h)^2}{v(\bm h)}.
\end{equation}
The notation $\overset{\mathrm A}{\sim}$ denotes the marginal distribution, after averaging over allelic effects and residual variation. Without direct effects, $\lambda_{\mathrm A}=0$ and we recover that the marginal distribution is a scaled central $\chi_1^2$. Direct effects shift the mean mismatch away from zero and yield a noncentral distribution; their influence depends on the directional projection $\bm c^\top\bm h$, not merely on the magnitude of $\bm c$.

For a fixed genetic architecture, the mismatch has mean $\mu_{\mathrm Q}(\bm h)=\bm\theta^\top\bm h$ and variance $\sigma_\varepsilon^2$. If $\sigma_\varepsilon^2>0$, then
\begin{equation}\label{eq:single-noncentral-Q}
 \frac{2\sigma_s^2}{\sigma_\varepsilon^2}Q(\bm h)
 \ \overset{\mathrm Q}{\sim}\ \chi_1^2\!\left(\lambda_{\mathrm Q}(\bm h)\right),\qquad
 \lambda_{\mathrm Q}(\bm h)=\frac{(\bm\theta^\top\bm h)^2}{\sigma_\varepsilon^2}.
\end{equation}

The chi-square representation also gives an exact expression for \emph{mean relative fitness}, which must not be confused with $\exp[-\mu_{\mathrm A}]$ or $\exp[-\mu_{\mathrm Q}]$. Marginally,
\begin{equation}\label{eq:single-meanomega-A}
 \overline\omega_{\mathrm A}(\bm h)\stackrel{\mathrm{def}}{=}\E_{a,\varepsilon}[\omega(\bm h)\mid\bm h  ]
 =\left(1+\frac{v(\bm h)}{\sigma_s^2}\right)^{-1/2}
 \exp\!\left[-\frac{(\bm c^\top\bm h)^2}{2(\sigma_s^2+v(\bm h))}\right].
\end{equation}
In particular, without direct effects,
\begin{equation}\label{eq:single-meanomega-A0}
 \overline\omega_{\mathrm A}(\bm h)
 =\left(1+\frac{\sigma_a^2g(\bm h)+\sigma_\varepsilon^2}{\sigma_s^2}\right)^{-1/2}.
\end{equation}
This expression is strictly decreasing in $g$ when $\sigma_a^2>0$, so squared genomic offset correctly ranks marginal mean relative fitness in the absence of direct effects, even though the relationship is nonlinear. It also illustrates the difference between predicting average log-fitness loss and predicting mean fitness: the former increases linearly with $g$, whereas the latter decreases according to an inverse square-root law.

Conditioning on a realized genetic architecture instead gives
\begin{equation}\label{eq:single-meanomega-Q}
 \overline\omega_{\mathrm Q}(\bm h;\bm\beta)
 \stackrel{\mathrm{def}}{=}\E_\varepsilon[\omega(\bm h)\mid\bm h,\bm\beta]
 =\left(1+\frac{\sigma_\varepsilon^2}{\sigma_s^2}\right)^{-1/2}
 \exp\!\left[-\frac{(\bm\theta^\top\bm h)^2}{2(\sigma_s^2+\sigma_\varepsilon^2)}\right].
\end{equation}
For $\bm c=\bm0$, this formula holds with $\bm\theta=\bm\beta$. Unlike the marginal mean, the quenched mean fitness generally depends on the orientation of the realized genetic effect vector relative to the environmental displacement. The identity $\E_a[\overline\omega_{\mathrm Q}]=\overline\omega_{\mathrm A}$ holds under the stated assumptions, but does not make the two predictions interchangeable.

\paragraph{Probability distribution function of relative fitness.}
Because $\omega=e^{-Q}$ is a decreasing transformation and $0<\omega \le1$, the distribution of fitness follows directly from that of the noncentral chi-square variable. For $0<t<1$, write $F_{\chi_1^2(\lambda)}$ for its cumulative distribution function. Then
\begin{align}
 \mathbb{P}_{\mathrm A}(\omega \le t\mid\bm h  )
 &=1-F_{\chi_1^2(\lambda_{\mathrm A})}
 \!\left(\frac{-2\sigma_s^2\log t}{v(\bm h)}\right),\label{eq:single-cdf-A}\\
 \mathbb{P}_{\mathrm Q}(\omega\le t\mid\bm h,\bm\beta)
 &=1-F_{\chi_1^2(\lambda_{\mathrm Q})}
 \!\left(\frac{-2\sigma_s^2\log t}{\sigma_\varepsilon^2}\right),\label{eq:single-cdf-Q}
\end{align}
for positive marginal and residual variances, respectively. The endpoint values follow by continuity. These formulas allow calculation of the probability that relative fitness falls below a biologically relevant threshold, rather than summarizing fitness only through its expectation. They concern relative fitness under the stabilizing-selection model; absolute survival or population persistence requires additional assumptions.

\subsection{Rank-one geometry and predictive accuracy across environments}\label{subsec:single-geometry}
For a realized architecture, the fitness matrix is
\begin{equation}\label{eq:single-rankone}
 M_B=\frac{\bm\theta\bm\theta^\top}{\sigma_s^2},\qquad
 \operatorname{rank}(M_B)\le1.
\end{equation}
Assume, as in Section~\ref{subsec:spectral}, that environmental displacements across the evaluation set follow $\bm h\sim\mathcal N(\bm0,\Sigma_x)$, with $\Sigma_x$ positive definite. When $\bm\theta\ne\bm0$ and $C_b\ne0$, the squared correlation between genomic offset and quenched mean log-fitness loss is
\begin{equation}\label{eq:single-R2}
 R_{\mathrm Q,1}^2
 =\frac{(\bm\theta^\top\Sigma_xC_b\Sigma_x\bm\theta)^2}
 {\tr[(C_b\Sigma_x)^2](\bm\theta^\top\Sigma_x\bm\theta)^2}.
\end{equation}
This expression holds with or without direct effects, through $\bm\theta=\bm\beta+\bm c$. Selection strength cancels from the correlation, although it changes the magnitude of fitness loss. If $\lambda_1\ge\cdots\ge\lambda_d\ge0$ are the eigenvalues of $A=\Sigma_x^{1/2}C_b\Sigma_x^{1/2}$, the spectral bounds reduce to
\begin{equation}\label{eq:single-spectral}
 \frac{\lambda_d^2}{\sum_i\lambda_i^2}
 \le R_{\mathrm Q,1}^2\le
 \frac{\lambda_1^2}{\sum_i\lambda_i^2}.
\end{equation}
Consequently, even a perfectly measured genomic offset may be a poor predictor of realized fitness if the GEA metric spans many equally important environmental directions but selection acts through a single latent trait.  
Strong prediction remains possible when genomic offset variation is concentrated along a dominant environmental gradient and the realized environmental response of the latent trait, represented by the vector $\Sigma_x^{1/2}\boldsymbol\beta$ in whitened coordinates, aligns with this gradient.
This provides a geometric route to predictive accuracy that is distinct from concentration through large $K_{\mathrm{eff}}$.

\section{Proofs of the main results}\label{sec:proofs}

Most mathematical arguments are based on quadratic forms and Gaussian moment calculations (see \citealp{Muirhead1982} for a useful reference).

\subsection{Proof of Proposition~\ref{prop:keff}: effective adaptive dimensionality}
Fix $\bm h$ and condition throughout on the GEA coefficients $\bm b$.
By Eq.~\eqref{eq:covU} and Gaussianity,
$\bm U(\bm h)\sim\mathcal N(\bm0,g(\bm h)\Sigma_a)$.
When $C=0$ and $\Sigma_\varepsilon=0$,

$$
 Q(\bm h)=\tfrac12\bm U(\bm h)^\top S\bm U(\bm h).
$$
Let $H=\Sigma_a^{1/2}S\Sigma_a^{1/2}\succeq0$ and
$\bm z\sim\mathcal N(\bm0,I_K)$. Then, in distribution,

$$
 Q(\bm h)=\frac{g(\bm h)}2\bm z^\top H\bm z.
$$
Diagonalizing $H$, with eigenvalues $\nu_1,\ldots,\nu_K\ge0$,
gives $\bm z^\top H\bm z\overset{d}=\sum_j\nu_j z_j^2$.
Since the $z_j$ are independent standard normals,

$$
 \E_a [Q]=\frac{g(\bm h)}2\sum_j\nu_j
 =\frac{g(\bm h)}2\tr(S\Sigma_a)=\gamma g(\bm h),
$$
and

$$
 \Var_a (Q)=\frac{g(\bm h)^2}{4}\sum_j\nu_j^2\Var(z_j^2)
 =\frac{g(\bm h)^2}{2}\tr(S\Sigma_a S\Sigma_a).
$$
Dividing by $\E_a [Q]^2$ yields Eq.~\eqref{eq:cv}.
\hfill$\square$

\subsection{Proof of Proposition~\ref{prop:matrix}: concentration of the genetic fitness matrix}
Under the conditional Gaussian assumptions,
$B\overset{d}=C_b^{1/2}Z\Sigma_a^{1/2}$, where $Z$ is a
$d\times K$ matrix with independent standard normal entries.
The representation also holds when either covariance is singular.
Using $H=\Sigma_a^{1/2}S\Sigma_a^{1/2}$ as in the preceding proof,

$$
 BSB^\top\overset{d}=C_b^{1/2}ZHZ^\top C_b^{1/2},
 \qquad \E_a[BSB^\top]=\tr(H)C_b=2\gamma C_b.
$$
By Isserlis' formula, for any indices $i,j,k,l$, we have

$$
 \E\big[(ZHZ^\top)_{ij}(ZHZ^\top)_{kl}\big] =\tr(H)^2\delta_{ij}\delta_{kl}
 +\tr(H^2)(\delta_{ik}\delta_{jl}+\delta_{il}\delta_{jk}).
$$
Subtracting the product of the means and contracting the indices
against $C_b$ gives
\begin{align*}
 \E_a [ \|BSB^\top-2\gamma C_b\|_F^2 ] \,  & =
 \E \left[\tr\!\left[
 (ZHZ^\top-\tr(H)I_d)C_b
 (ZHZ^\top-\tr(H)I_d)C_b\right]\right]\\
 &=\tr(H^2)\bigl( \tr(C_b^2)+[\tr(C_b)]^2\bigr).
\end{align*}
Since $\tr(H)=2\gamma$ and
$\tr(H^2)=\tr(S\Sigma_a S\Sigma_a)
=(2\gamma)^2/K_{\mathrm{eff}}$, division by $(2\gamma)^2$
proves Eq.~\eqref{eq:matrixconcentration}.
The same calculation applied to $U=\Sigma_x^{1/2}B$, whose
row covariance is $A$, yields
\begin{equation}\label{eq:proof-genetic-whitened}
 \E_a \left[\|USU^\top-2\gamma A\|_F^2 \right]
 =(2\gamma)^2\|A\|_F^2
 \frac{1+r_{\mathrm{eff}}(A)}{K_{\mathrm{eff}}}.
\end{equation}
\hfill$\square$

\subsection{Quadratic-form correlations and the quenched prediction coefficient}
Let $\bm h\sim\mathcal N(\bm0,\Sigma_x)$, with $\Sigma_x\succ0$.
For symmetric matrices $P,N$, consider the quadratic forms
$\bm h^\top P\bm h$ and $\bm h^\top N\bm h$.
Writing $\bm h=\Sigma_x^{1/2}\bm z$ with
$\bm z\sim\mathcal N(\bm0,I_d)$ and applying Isserlis' identity,
$\E[z_i z_j z_k z_l]=\delta_{ij}\delta_{kl}
+\delta_{ik}\delta_{jl}+\delta_{il}\delta_{jk}$, gives

$$
 \E_{\bm h} [\bm h^\top P\bm h ] =\tr(P\Sigma_x),
$$
and

$$
 \E_{\bm h}[(\bm h^\top P\bm h)(\bm h^\top N\bm h)] =\tr(P\Sigma_x)\tr(N\Sigma_x)
                  +2\tr(P\Sigma_xN\Sigma_x).
$$
Therefore
\begin{equation}\label{eq:proof-quadratic-cov}
 \Cov_{\bm h}(\bm h^\top P\bm h,\bm h^\top N\bm h)=2\tr(P\Sigma_xN\Sigma_x),\qquad
 \Var_{\bm h}(\bm h^\top P\bm h)=2\tr[(P\Sigma_x)^2].
\end{equation}
For any nonzero positive semidefinite fitness matrix $M$ and $C_b$, the constants
and positive scalar factors in the two quadratic responses cancel
from their Pearson correlation. Hence
\begin{equation}\label{eq:proof-general-r2}
 R^2(M)=
 \frac{[\tr(C_b\Sigma_xM\Sigma_x)]^2}
 {\tr[(C_b\Sigma_x)^2]\,\tr[(M\Sigma_x)^2]}.
\end{equation}
In the quenched model, $M=M_B=(B+C)S(B+C)^\top$, and thus
\begin{equation}\label{eq:proof-r2-quenched}
 R_{\mathrm Q}^2=
 \frac{[\tr(C_b\Sigma_xM_B\Sigma_x)]^2}
 {\tr[(C_b\Sigma_x)^2]\,\tr[(M_B\Sigma_x)^2]}.
\end{equation}
Equivalently, in whitened coordinates, with
$A=\Sigma_x^{1/2}C_b\Sigma_x^{1/2}$,
$F=\Sigma_x^{1/2}M_B\Sigma_x^{1/2}$ and
$J=\Sigma_x^{1/2}CSC^\top\Sigma_x^{1/2}$,
\begin{equation}\label{eq:proof-both-r2}
 R_{\mathrm Q}^2=\frac{\langle A,F\rangle_F^2}
 {\|A\|_F^2\|F\|_F^2},
\end{equation}
All correlations above are defined when the relevant quadratic
forms have nonzero variance. \hfill$\square$

\subsection{Proof of Proposition~\ref{prop:spectral}: spectral bounds}
By Eq.~\eqref{eq:proof-both-r2},
$R_{\mathrm Q}^2=\langle A,F\rangle_F^2/
(\|A\|_F^2\|F\|_F^2)$.
Let $r=\operatorname{rank}(F)\le\min(K,d)$ and let
$\mu_1\ge\cdots\ge\mu_r>0$ be its nonzero eigenvalues.
Von Neumann's trace inequality and Cauchy-Schwarz imply

$$
 \tr(AF)\le\sum_{i=1}^r\lambda_i\mu_i
 \le\left(\sum_{i=1}^r\lambda_i^2\right)^{1/2}
          \left(\sum_{i=1}^r\mu_i^2\right)^{1/2}
 =\left(\sum_{i=1}^r\lambda_i^2\right)^{1/2}\|F\|_F.
$$
This proves the sharper rank-dependent upper bound
\[
 R_{\mathrm Q}^2\le
 \frac{\sum_{i=1}^r\lambda_i^2}{\sum_{i=1}^d\lambda_i^2}
 \le\frac{\sum_{i=1}^{\min(K,d)}\lambda_i^2}
 {\sum_{i=1}^d\lambda_i^2}.
\]
For the lower bound, $A\succeq\lambda_d I_d$ and $F\succeq0$ give
\[
 \tr(AF)\ge\lambda_d\tr F\ge\lambda_d\|F\|_F,
\]
where the second inequality follows from the nonnegative eigenvalues
of $F$. Therefore
$R_{\mathrm Q}^2\ge\lambda_d^2/\sum_i\lambda_i^2$.
The lower bound is attained by $F=vv^\top$, where $v$ is a unit
eigenvector for $\lambda_d$. The upper bound for rank at most $K$
is attained by choosing $F$ proportional to the spectral truncation
$\sum_{i=1}^{\min(K,d)}\lambda_i v_iv_i^\top$ (when this matrix is
nonzero). If $\lambda_d=0$, the lower bound is zero and is attained
by a rank-one matrix supported in $\ker A$.
\hfill$\square$

\subsection{Proof of Proposition~\ref{prop:direct}: Reliability of realized predictions with direct environmental effects}
By Eq.~\eqref{eq:FX},
$F=USU^\top+J+X$ and $X=USD^\top+DSU^\top$.
The conditional Gaussian distribution of $U$ satisfies
\begin{equation}\label{eq:proof-U-moments}
 \E_a[U_{ip}]=0,\qquad
 \E_a[U_{ip}U_{jq}]=A_{ij}(\Sigma_a)_{pq},
 \qquad \E_a[USU^\top]=2\gamma A.
\end{equation}
Since $X$ is linear in $U$, $\E_a[X]=0$, hence
$\E_a[F]=2\gamma A+J$.

To compute the interaction term moment, expand the Frobenius norm:
\begin{align*}
 \|X\|_F^2
 &=2\tr(USD^\top DSU^\top)
   +2\tr(USD^\top USD^\top).
\end{align*}
For the first term, Eq.~\eqref{eq:proof-U-moments} implies
$\E_a[U^\top U]=\tr(A)\Sigma_a$, so that
\begin{align*}
 \E_a[\tr(USD^\top DSU^\top)] &=\tr\!\left(\E_a[U^\top U]SD^\top DS\right)\\
 &=\tr(A)\tr(SD^\top DS\Sigma_a).
\end{align*}
For the second term, expanding the trace and contracting the
Gaussian second moments gives
\begin{align*}
 \E_a[\tr(USD^\top USD^\top)] \,
 &=\sum_{i,j,p,q,r,s}
 \E_a[U_{ip}U_{jr}]S_{pq}D_{jq}S_{rs}D_{is}\\
 &=\sum_{i,j,p,q,r,s}
 A_{ij}(\Sigma_a)_{pr}S_{pq}D_{jq}S_{rs}D_{is}\\
 &=\tr(ADS\Sigma_a SD^\top).
\end{align*}
This establishes Eq.~\eqref{eq:crossvariance}.

Next, $F-2\gamma A=(USU^\top-2\gamma A)+J+X$.
The terms involving $J$ have zero expected cross products,
because $\E_a \left[USU^\top-2\gamma A \right]=\E_a[X]=0$.
Moreover,

$$
 \E_a\left[\langle USU^\top-2\gamma A,X\rangle_F \right]=0,
$$
since the integrand is an odd polynomial in the jointly centered
Gaussian entries of $U$. Therefore
\begin{align*}
 \E_a\left[\|F-2\gamma A\|_F^2\right]
 &=\E_a\left[\|USU^\top-2\gamma A\|_F^2\right]
   +\|J\|_F^2+\E_a[\|X\|_F^2] \\
 &=(2\gamma)^2\|A\|_F^2
 \frac{1+r_{\mathrm{eff}}(A)}{K_{\mathrm{eff}}}
   +\|J\|_F^2+\E_a[\|X\|_F^2] \, ,
\end{align*}
where the last equality uses Eq.~\eqref{eq:proof-genetic-whitened}.
Dividing by $(2\gamma)^2$ proves Eq.~\eqref{eq:directmoment}.

For any $0<t<1$, Markov's inequality and
Eq.~\eqref{eq:directmoment} imply
\[
 \mathbb P_a\!\left(
 \left\|\frac{F}{2\gamma}-A\right\|_F>t\|A\|_F
 \right)\le\frac{\eta_{\mathrm{tot}}}{t^2}.
\]
On the complementary event,
$\|F/(2\gamma)\|_F\ge(1-t)\|A\|_F>0$.
Let $\Pi_A$ denote the Frobenius orthogonal projection onto
$\operatorname{span}\{A\}$. By Eq.~\eqref{eq:r2},
\begin{align*}
 1-R_{\mathrm Q}^2
 &=\frac{\|(I-\Pi_A)F/(2\gamma)\|_F^2}
 {\|F/(2\gamma)\|_F^2}\\
 &=\frac{\|(I-\Pi_A)(F/(2\gamma)-A)\|_F^2}
 {\|F/(2\gamma)\|_F^2}
 \le\frac{t^2}{(1-t)^2}.
\end{align*}
This proves Eq.~\eqref{eq:directprob}. If
$\eta_{\mathrm{tot}}\to0$, then
$\|F/(2\gamma)-A\|_F/\|A\|_F\to0$ in probability,
and the same bound yields $R_{\mathrm Q}^2\to1$ in probability.
\hfill$\square$

\subsection{Proof of Corollary~\ref{cor:alignment}: Geometric alignment of direct environmental effects}
Assume $J\ne0$; the case $J=0$ follows by setting $d_D=0$.
Let $\Pi_A$ again be Frobenius projection onto the span of $A$.
Decompose
\[
 J=\rho A+J_\perp,\qquad
 \rho=\frac{\langle A,J\rangle_F}{\|A\|_F^2},\qquad
 \langle A,J_\perp\rangle_F=0.
\]
By Pythagoras and Eq.~\eqref{eq:rhoD},
\[
 \frac{\rho}{2\gamma}=d_D\rho_D,
 \qquad
 \frac{\|J_\perp\|_F}{2\gamma\|A\|_F}
 =d_D\sqrt{1-\rho_D^2}.
\]
By Eq.~\eqref{eq:proof-genetic-whitened} and the vanishing
odd Gaussian moments,

$$
 \E_a\left[\|USU^\top-2\gamma A+X\|_F^2\right]
 =4\gamma^2 \eta \|A\|_F^2.
$$
Consequently, Markov's inequality gives

$$
 \mathbb P_a\!\left(
 \|USU^\top-2\gamma A+X\|_F
 \le 2\gamma t\|A\|_F\right)
 \ge1-\frac{\eta}{t^2}.
$$
On this event, using Eq.~\eqref{eq:FX},

$$
 \frac{F}{2\gamma}
 =(1+d_D\rho_D)A+\frac{J_\perp}{2\gamma}
  +\frac{USU^\top-2\gamma A+X}{2\gamma}.
$$
By Eq.~\eqref{eq:orthdirect} and the triangle inequality,

$$
 \left\|\frac{J_\perp+USU^\top-2\gamma A+X}{2\gamma}\right\|_F
 \le b\|A\|_F,
 \qquad b=d_D\sqrt{1-\rho_D^2}+t.
$$
Writing $a=1+d_D\rho_D$ as in the corollary and assuming $b<a$,
the reverse triangle inequality gives
$\|F/(2\gamma)\|_F\ge(a-b)\|A\|_F>0$.
Since $(I-\Pi_A)A=0$,

$$
 \left\|(I-\Pi_A)F/(2\gamma)\right\|_F\le b\|A\|_F.
$$
Thus the squared cosine identity yields

$$
 1-R_{\mathrm Q}^2
 =\frac{\|(I-\Pi_A)F/(2\gamma)\|_F^2}
 {\|F/(2\gamma)\|_F^2}
 \le\frac{b^2}{(a-b)^2},
$$
which proves Eq.~\eqref{eq:alignmentbound}.
For the limiting assertion, $a=1+d_D\rho_D\ge1$.
If $d_D\sqrt{1-\rho_D^2}\to0$ and $\eta\to0$, choose
$t\to0$ with $\eta/t^2\to0$; then $b\to0$ and the probability
bound tends to one, so $R_{\mathrm Q}^2\to1$ in probability.
\hfill$\square$

\section{Discussion}\label{sec:discussion}

A central conclusion of this study is that accurate genomic offset (GO) predictions do not uniquely identify the underlying adaptive architecture. Similar predictive performance can arise from substantially different biological architectures, so a strong association between GO and fitness loss does not, by itself, reveal whether adaptation is effectively high- or low-dimensional.

This conclusion rests on the distinction between marginal (annealed) and realized (quenched) predictions. Under the assumed distribution of allelic effects and stabilizing selection, and in the absence of direct environmental effects, squared GO is proportional to the marginal expectation of negative log-relative fitness \citep{Gain2023,Capblancq2025}. For a particular genetic architecture, however, the fitness-loss matrix need not be proportional to the GEA metric. Two complementary mechanisms can nevertheless support accurate predictions. First, when many effectively independent adaptive traits contribute to fitness, the genetically mediated response concentrates around its marginal expectation. This depends on effective adaptive dimensionality, not simply on the number of loci. Second, even with few adaptive traits, prediction can be accurate when the complete fitness-loss matrix is geometrically aligned with the environmentally whitened GEA metric. This second mechanism includes direct environmental effects and their interactions with genetically mediated responses.

These results offer a possible interpretation of differences in predictive performance reported in evolutionary simulations [\citealp{Laruson2022,Gain2023}, Lind and Lotterhos 2024,\citealp{Capblancq2025}]. Beyond the number of selected loci and the quality of GEA estimation, such differences may reflect effective adaptive dimensionality or the alignment of fitness-relevant environmental gradients with those captured by GO. This interpretation remains to be tested directly: simulations retaining allelic effects and selection matrices would allow the concentration and geometric predictions to be evaluated without inferring latent quantities from offset statistics alone.

Geometric alignment also has a direct biological interpretation. A small number of adaptive traits can suffice for accurate prediction of genetically mediated fitness loss when their effects on fitness follow the dominant environmental gradients captured by GO. Likewise, direct environmental responses can contribute substantially to fitness loss without impairing prediction if they preserve the geometry represented by the GEA metric; contributions along poorly aligned gradients can instead weaken prediction. Yet direct-effect alignment alone is not sufficient for a realized architecture. With centered allelic effects, genetic--direct interaction terms vanish on averaging over architectures but generally persist for any particular architecture. These interactions can reinforce or compensate genetically mediated mismatch and change the geometry of the complete fitness-loss matrix, whose overall alignment with the GEA metric ultimately determines predictive accuracy.

The theory also identifies structural limits to prediction, even when GEA coefficients are accurately estimated. With few effectively independent adaptive traits, realized responses may differ markedly from their marginal expectation. In particular, a low-rank fitness response need not reproduce a GEA metric spread across several environmental directions of comparable importance. Poor alignment, including contributions from direct environmental effects along gradients weakly represented by GEA, can further reduce predictive accuracy. Thus, neither a large number of loci nor accurate GEA estimation alone guarantees reliable fitness predictions.

A separate limitation concerns observed fitness. Even if GO predicts the conditional mean negative log-relative fitness for a realized architecture (high $R_{\mathrm Q}^2$), residual variation in adaptive traits reduces the proportion of observed variation explained by GO, as quantified by $R_{\mathrm{obs}}^2$ in Section~3.3. A weak empirical association can therefore reflect individual variation rather than poor geometric alignment. Common-garden results \citep{Rhone2020,Fitzpatrick2021,Lachmuth2023} are consistent with GO capturing components of adaptive mismatch, but the strength of this association need not be universal. Moreover, the present framework concerns relative fitness under specified stabilizing-selection assumptions, not absolute fitness across environments with different productivity. Missing environmental drivers, nonequilibrium adaptation, and weak or absent local fitness trade-offs remain important limitations \citep{Hereford2009,Kawecki2004,Lotterhos2023,Lotterhos2024,Ahrens2025}.

In summary, GO is best interpreted as a genetically informed measure of adaptive mismatch, rather than a direct predictor of realized fitness or any particular measured phenotype. Its predictive relevance depends on the latent adaptive architecture and on the environmental geometry of fitness loss. When the concentration or alignment conditions identified here hold, GO can nevertheless rank environmental displacements according to their potential fitness consequences.

\vspace{1cm}
\paragraph{Data availability statement.} Not applicable to this work.

\vspace{0.5cm}

\paragraph{Acknowledgments.} This work received support from the French National Research Agency, project PEG2 (grant number ANR-22-CE45-0033) and Afradapt (grant number ANR-22-CE32-0008).


\vspace{1cm}

\begin{thebibliography}{}

\bibitem[Ahrens {\it et al.} 2025]{Ahrens2025} Ahrens, C. W., Rymer, P. D., Miller, A. D. (2025). Genetic offset and vulnerability modelling under climate change scenarios: common misinterpretations and violations of evolutionary principles. Evolution, 80(1), 15-27.
doi: 10.1093/evolut/qpaf216

\bibitem[Arnold {\it et al.} 2001]{Arnold2001} Arnold, S. J., Pfrender, M. E., Jones, A. G. (2001). The adaptive landscape as a conceptual bridge between micro- and macroevolution. Genetica, 112--113, 9--32. doi: 10.1023/A:1013373907708.

\bibitem[Capblancq {\it et al.} 2025]{Capblancq2025} Capblancq, T., Cubry, P., Tauzin, A., Vigouroux, Y., Fran\c cois, O. (2025). Linking genomic offset statistics to the shape of selection gradients. The American Naturalist, 207(3), 356-367. doi: 10.1086/739079.

\bibitem[Fitzpatrick and Keller 2015]{Fitzpatrick2015}  Fitzpatrick, M. C.,  Keller, S. R. (2015). Ecological genomics meets community-level modelling of biodiversity: Mapping the genomic landscape of current and future environmental adaptation. Ecology Letters, 18(1), 1-16. doi: 10.1111/ele.12376.

\bibitem[Fitzpatrick {\it et al.} 2021]{Fitzpatrick2021} Fitzpatrick, M. C., Chhatre, V. E., Soolanayakanahally, R. Y.,  Keller, S. R. (2021). Experimental support for genomic prediction of climate maladaptation using the machine learning approach Gradient Forests. Molecular Ecology Resources, 21(8), 2749-2765. doi: 10.1111/1755-0998.13374.

\bibitem[Forester {\it et al.} 2018]{Forester2018} Forester, B. R., Lasky, J. R., Wagner, H. H., Urban, D. L. (2018). Comparing methods for detecting multilocus adaptation with multivariate genotype-environment associations. Molecular Ecology, 27(9), 2215-2233.  doi: 10.1111/mec.14584.

\bibitem[Gain  {\it et al.} 2023]{Gain2023} Gain, C., Rhon\'e, B., Cubry, P., Salazar, I., Forbes, F., Vigouroux, Y., {\it et al.} (2023). A quantitative theory for genomic offset statistics. Molecular Biology and Evolution, 40(6), msad140. doi: 10.1093/molbev/msad140

\bibitem[Hansen and Houle 2008]{HansenHoule2008} Hansen, T. F., Houle, D. (2008). Measuring and comparing evolvability and constraint in multivariate characters. Journal of Evolutionary Biology, 21(5), 1201-1219. doi: 10.1111/j.1420-9101.2008.01573.x.

\bibitem[Hereford 2009]{Hereford2009} Hereford, J. (2009). A quantitative survey of local adaptation and fitness trade-offs. The American Naturalist, 173(5), 579-588. doi: 10.1086/597611.

\bibitem[Kawecki and Ebert 2004]{Kawecki2004}  Kawecki, T. J.,  Ebert, D. (2004). Conceptual issues in local adaptation. Ecology letters, 7(12), 1225-1241. doi: 10.1111/j.1461-0248.2004.00684.x.

\bibitem[Kirkpatrick 2009]{Kirkpatrick2009} Kirkpatrick, M. (2009).
Patterns of quantitative genetic variation in multiple dimensions. Genetica, 136, 271-284. doi: 10.1007/s10709-008-9302-6.

\bibitem[Lachmuth {\it et al.} 2023]{Lachmuth2023}  Lachmuth, S., Capblancq, T., Prakash, A., Keller, S. R., Fitzpatrick, M. C. (2023). Novel genomic offset metrics integrate local adaptation into habitat suitability forecasts and inform assisted migration. Ecological Monographs, e1593. doi: doi.org/10.1002/ecm.1593.

\bibitem[Lande 1979]{Lande1979} Lande, R. (1979). Quantitative genetic analysis of multivariate evolution, applied to brain:body size allometry. Evolution, 33(1), 402-416. doi: 10.1111/j.1558-5646.1979.tb04694.x.

\bibitem[Lande and Arnold 1983]{Lande1983} Lande, R., Arnold, S. J. (1983). The measurement of selection on correlated characters. Evolution, 37(6), 1210-1226. doi: 10.2307/2408842.

\bibitem[L\`aruson {\it et al.} 2022]{Laruson2022} L\`aruson, \`A. J., Fitzpatrick, M. C., Keller, S. R., Haller, B. C., Lotterhos, K. E. (2022). Seeing the Forest for the trees: Assessing genetic offset predictions with Gradient Forest. Evolutionary Applications, 15(3), 403-416. doi: 10.1111/eva.13354


\bibitem[Lasky {\it et al.} 2023]{Lasky2023} Lasky, J. R., Josephs, E. B.,  Morris, G. P. (2023). Genotype–environment associations to reveal the molecular basis of environmental adaptation. The Plant Cell, 35(1), 125-138. doi: 10.1093/plcell/koac267.

\bibitem[Lind and Lotterhos 2024]{Lind2024} Lind , B. M.,  Lotterhos, K. E. (2024). The accuracy of predicting maladaptation to new environments with genomic data. Molecular Ecology Resources, 25, e14008. doi: 10.1111/1755-0998.14008.

\bibitem[Lotterhos 2023]{Lotterhos2023} Lotterhos, K. E. (2023). The paradox of adaptive trait clines with nonclinal patterns in the underlying genes. Proceedings of the National Academy of Sciences, 120(12), e2220313120. doi: 10.1073/pnas.2220313120.

\bibitem[Lotterhos 2024]{Lotterhos2024} Lotterhos, K. E. (2024). Interpretation issues with ``genomic vulnerability'' arise from conceptual issues in local adaptation and maladaptation. Evolution Letters, 8(3), 331-339. doi: 10.1093/evlett/qrae004.

\bibitem[Muirhead 1982]{Muirhead1982} Muirhead, R. J. (1982). Aspects of Multivariate Statistical Theory. John Wiley \& Sons.

\bibitem[Pigliucci 2001]{Pigliucci2001} Pigliucci, M. (2001). Phenotypic Plasticity: Beyond Nature and Nurture. Johns Hopkins University Press.

\bibitem[Rellstab {\it et al.} 2016]{Rellstab2016}  Rellstab, C., Zoller, S., Walthert, L.,  Bod\'en\`es, C., Lesur, I., Pluess, A. R., (2016). Signatures of local adaptation in candidate genes of oaks (Quercus spp.) with respect to present and future climatic conditions. Molecular Ecology, 25(23), 5907-5924. doi: 10.1111/mec.13889.

\bibitem[Rellstab {\it et al.} 2021]{Rellstab2021}  Rellstab, C., Dauphin, B., Exposito-Alonso, M. (2021). Prospects and limitations of genomic offset in conservation management. Evolutionary Applications, 14(5), 1202-1212. doi: 10.1111/eva.13205.

\bibitem[Rhon\'e {\it et al.} 2020]{Rhone2020}  Rhon\'e, B., Defrance, D., Berthouly-Salazar, C., Mariac, C., Cubry, P., Couderc, M., {\it et al.} (2020). Pearl millet genomic vulnerability to climate change in West Africa highlights the need for regional collaboration. Nature Communications, 11(1), 5274. doi:10.1038/s41467-020-19066-4

\bibitem[Walsh and Lynch 2018]{WalshLynch2018} Walsh, B., Lynch, M. (2018). Evolution and Selection of Quantitative Traits. Oxford University Press.

\bibitem[Yeaman 2015]{Yeaman2015} Yeaman, S. (2015). Local adaptation by alleles of small effect. The American Naturalist, 186(S1), S74--S89. doi: 10.1086/682405.




\end{thebibliography}
\end{document}